# Structure of chiral single-walled carbon nanotubes under hydrostatic pressure

Ali Nasir Imtani and V. K. Jindal[1]

Department of Physics, Panjab University, Chandigarh-160014, India

We investigate the structural parameters, i.e. bond lengths and bond angles of chiral tubes of various chiralities. The procedure used is based on helical and rotational symmetries and Tersoff potential. The results indicate that at ambient condition, there are equal bond lengths and three unequal bond angles in the structure of chiral tubes. The bond length depends much more on the chirality and very slightly on the tube radius. Length of the tubes does not play very significant role on bond length and bond angles. These C-C bonds were recalculated under hydrostatic pressure. The bond length compresses with pressure while the bond angles remain practically unchanged. We also carry out analysis regarding the cross sectional shape of chiral tubes and its pressure dependence. It is found that at some pressures, transition from circular to oval cross section takes place. The transition pressure is found to strongly depend on the radius and chirality of tube. At this transition, corresponding to given elliptical cross section, the bond length for all chiral tubes is identical. This behavior of bond length is different from achiral tubes.

## I- Introduction

Interest in carbon nanotubes (CNTs) continues to grow since their first discovery [1]. CNTs possess many novel and unique properties including structural perfection, low density, high stiffness and strength and excellent electric properties and bio-compatibility. As a result, CNTs may have a wide range of technological applications. Single-wall carbon nanotubes(SWNTs) are characterized by two integers (n,m) defining the rolling vector of graphite[2]. The strong similarity of the chemistry of carbon nanotubes to graphite allows theoretical analysis to be done based on empirical methodologies imported from studies on graphite. The curvature of the tubes, however, disturbs the chemistry and causes deviation from the graphite based description, especially for small radii tubes. The structure of SWNTs is qualitatively well known through the simple construction of rolling a perfect graphite sheet, where only one parameter has been considered; the lattice parameter or the bond length. It is very difficult to obtain the direct experimental information for the structure, although a lot of theoretical information is available [3-5]. There have been many researches on the variation of the structural parameters with tube radius for armchair and zigzag SWNTs [6-11].

---

[1] Author with whom correspondence be made, e-mail: jindal@pu.ac.in

One way to study the structural and mechanical properties of carbon nanotubes is by applying hydrostatic pressure. A number of high-pressure experiments have been carried out on individual and bundles SWNTs [12-21], showing pressure induced structural transition.

So far theoretical studies on single-wall nanotubes under pressure have mostly focused on armchair and zigzag tubes [10-11,22-23], which have a high symmetrical radial atomic structure and a short axial period. These studies suggest that the tubes undergo transition at critical pressures resulting in modification of nanotubes cross section from circular to oval shape. Calculations on many properties of single-wall nanotubes like Young's modulus, bulk and structural properties and thermodynamical properties have been reported [24-27] using one bond length equivalent to that of graphite or modified value. The electronic band structure of semiconducting and metallic nanotubes under uniaxial strain is calculated using tight-binding approximations [27]. They used unstrained bond length equal to 1.44 Å for all the tubes.

In order to have an insight into such pressure induced structural transformation, it is preferable to have a detailed study based on established model potential. Further, it is necessary to understand the behavior of bond lengths under pressure, for which no detailed study exist.

This paper reports the results of the structure of chiral single-wall nanotubes $(n,m)$ with different radii and different chirality (i.e., chirality=$m/n$). We choose chiral tubes of varying chirality indicated by angles $4.175004°$ ($m/n=0.1$), $8.94876°$($m/n=0.2$, $13.892°$($m/n=0.33$), $19.1066°$($m/n=0.5$) and $26.3295°$($m/n=0.8$). The number of atoms is equal to $4(n^2+nm+m^2)/\gcd_R$ for $(n,m)$ tubes and different for different radii and chirality, where $\gcd_R$ is the greatest common divisor of integers $((2n+m),(2m+n))$. We have taken all tubes of nearly the same length, close to 125 Å as for this length, the values of bond lengths for achiarl tubes had saturated to constant value and therefore were treated as "long" tubes. The actual length is determined by N, the number of unit cell length steps that are necessary to give length closest to 120 A. For various chiral tubes chosen by us, this N equals 8, 5, 8, 11, 4.

## II. Helical and rotational symmetries

In general, there are three bond lengths $b_1$, $b_2$ and $b_3$ and three bond angles $\alpha$, $\beta$ and $\gamma$ in the structure of chiral (n,m) tubes, where $n \neq m$. A typical chiral tube is shown in Fig.1. We observe that all the bond lengths make an angle with the tube axis. Because the directions of two bond lengths $b_1$ and $b_3$ with the tube axis are equivalent, we treat these bond lengths equal. In our previously works [10, 11], we found that, at ambient pressure, SWNTs have a circular cross sectional shape.

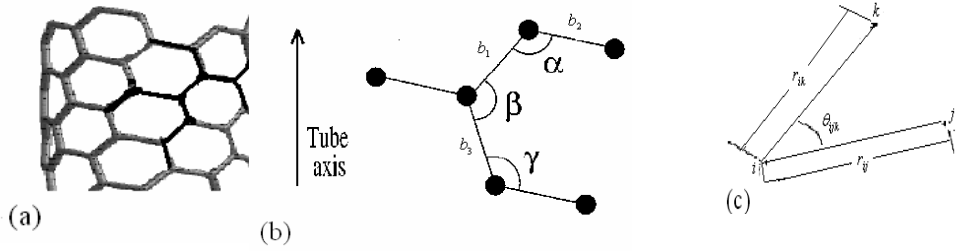

Figure 1: (a) A schematic side view of chiral SWNT and (b) a part of chiral SWNT indicating three types of the C-C bond lengths, these are labeled as $b_1$, $b_2$ and $b_3$, and three bond angles $\beta$, $\gamma$ and $\alpha$. (c) Carbon atoms $i$, $j$ and $k$, the corresponding bond lengths $i-j$ and $i-k$ and bond angles $\theta_{ijk}$.

The helical and rotational symmetries [28] are used to construct the chiral (n,m) SWNT's (with $0^o < \theta < 30^o$ and $n > m$). This is done by first mapping the two atoms in the [0,0] unit cell to the surface of cylindrical shape. The first atom is mapped to an arbitrary point on the cylindrical surface {e.g., $(R,0,0)$}, where $R$ is the tube radius in terms of bond lengths $b_1$ and $b_2$. The position of the second atom is found by rotating this point by $\phi = \pi \dfrac{n+m}{n^2+nm+m^2}$ around the cylinder axis in conjunction with translating by $h_t = \dfrac{1}{2}b_1 \dfrac{(n-m)}{\sqrt{n^2+nm+m^2}}$ along this axis. The cylinder axis must coincide with $\vec{C}_{gcd}$ axis for the tubule. These first two atoms can be used to locate $2(gcd-1)$ additional atoms on the cylindrical surface by $(gcd-1)$ successive $2\pi/gcd$ rotations about the cylinder axis, where gcd the greatest common divisor of (n,m). Altogether, these atoms complete the specification of the helical motif which corresponds to an area on the cylindrical surface. This helical can then be used to tile the reminder of the tubule by repeated operation of a single screw operation $S(h,\alpha_h)$ represent a translation $h(n,m) = \dfrac{3}{2}b_2 \dfrac{gcd}{\sqrt{n^2+nm+m^2}}$ along the cylinder axis and rotation $\alpha_h = \pi \dfrac{[n(2p_1+p_2)+m(2p_2+p_1)]}{n^2+nm+m^2}$ about this axis, where $p_1$ and $p_2$ are integers. If we apply the full helical motif, then the entire structure of chiral SWNT is generated. This helical motif provides atomic position of all atoms in terms of bond lengths. The bond lengths are determined by minimization of the energy of the tube, assuming atoms interact via Tersoff potential [10,29].

**III- Results and Discussion**

**(A) Effect of tube radius and chirality on the structure**

We start the chiral nanotube structure assuming two different bond lengths and three different angles and use Tersoff potential to find out the minimum energy configuration by varying systematically these bond lengths and bond angles. In contrast with the results of achiral tubes [10,11], we have found that there are equal bond lengths ($b_1 = b_2 = b_3 = b$) and three unequal bond angles in the structure of chiral tubes. In Table I, we present the results of our calculations of one bond length ($b$) and three bond angles for chiral tubes obtained by the procedure outline in our previous works on armchair SWNTs[10]. We also notice that all chiral tubes have a bond length significantly larger than that graphite bond length and does not approach to graphitic value for even larger radius tube, as large as 123.324 Å. For a fixed tube length, there are two variables which are affecting the structural parameters of chiral tubes. These variables are the tube radius and chiral angle (or chirality). We can also see that, for the same chirality, the bond length for different radii tubes has nearly the same value except for very small radii tubes.

Table I: Radius (Å), bond length, bond angles and energy (eV/atom) for chiral SWNTs of various chiral angles.

| (a) Chiral angle $\theta$ =4.1750° | | | | | | |
|---|---|---|---|---|---|---|
| SWNT | Radius | $b$ (Å) | $\alpha$ | $\beta$ | $\gamma$ | Energy |
| (10,1) | 4.17348 | 1.437 | 117.312 | 123.715 | 116.367 | -7.05703 |
| (20,2) | 8.33593 | 1.4351 | 117.221 | 125.639 | 116.368 | -7.05115 |
| (30,3) | 12.5030 | 1.435 | 117.205 | 126.002 | 116.368 | -7.05415 |
| (40,4) | 16.6649 | 1.4345 | 117.199 | 126.130 | 116.370 | -7.05733 |
| (50,5) | 20.8311 | 1.4345 | 117.196 | 126.190 | 116.381 | -7.05815 |
| (100,10) | 83.3533 | 1.4345 | 117.192 | 126.289 | 116.367 | -7.05911 |

| (b) Chiral angle θ =8.9487° | | | | | | |
|---|---|---|---|---|---|---|
| SWNT | Radius | $b$ (Å) | $\alpha$ | $\beta$ | $\gamma$ | *Energy* |
| (5,1) | 2.2516 | 1.467 | 116.412 | 121.351 | 113.236 | -6.64133 |
| (10,2) | 4.4663 | 1.455 | 116.081 | 127.945 | 113.462 | -6.64710 |
| (15,3) | 6.6995 | 1.455 | 116.020 | 129.263 | 113.505 | -6.65399 |
| (25,5) | 11.1659 | 1.455 | 115.988 | 129.952 | 113.527 | -6.66236 |
| (30,6) | 13.399 | 1.455 | 115.983 | 130.071 | 113.531 | -6.66321 |
| (40,8) | 17.865 | 1.455 | 115.977 | 130.190 | 113.535 | -6.66356 |
| (240,48) | 107.192 | 1.455 | 115.971 | 130.339 | 113.540 | -6.66532 |

| (c) Chiral angle θ=13.8922° | | | | | | |
|---|---|---|---|---|---|---|
| SWNT | Radius | $b$ (Å) | $\alpha$ | $\beta$ | $\gamma$ | *Energy* |
| (9,3) | 4.3891 | 1.472 | 116.017 | 130.075 | 111.205 | -6.38511 |
| (12,4) | 5.8522 | 1.472 | 115.971 | 131.125 | 111.304 | -6.38723 |
| (21,7) | 10.241 | 1.472 | 115.931 | 132.057 | 111.391 | -6.39001 |
| (36.12) | 17.556 | 1.471 | 115.918 | 132.354 | 111.418 | -6.43711 |
| (252,84) | 122.813 | 1.471 | 115.911 | 132.507 | 111.432 | -6.44101 |

| (d) Chiral angle θ=19.1066° | | | | | | |
|---|---|---|---|---|---|---|
| SWNT | Radius | $b$ (Å) | $\alpha$ | $\beta$ | $\gamma$ | *Energy* |
| (4,2) | 2.1573 | 1.479 | 117.467 | 123.333 | 109.510 | -6.4642 |
| (6,3) | 3.2185 | 1.471 | 117.268 | 127.704 | 110.459 | -6.4650 |
| (10,5) | 5.3533 | 1.468 | 117.165 | 130.102 | 110.956 | -6.4656 |
| (16,8) | 8.5595 | 1.467 | 117.130 | 130.953 | 111.128 | -6.4716 |
| (20,10) | 10.706 | 1.467 | 117.122 | 131.152 | 111.167 | -6.4773 |
| (30,15) | 16.049 | 1.467 | 117.114 | 131.349 | 111.207 | -6.4842 |
| (210,105) | 112.343 | 1.467 | 117.108 | 131.504 | 111.238 | -6.4892 |

| (e) Chiral angle=26.3295° | | | | | | |
|---|---|---|---|---|---|---|
| SWNT | Radius | $b$ (Å) | $\alpha$ | $\beta$ | $\gamma$ | Energy |
| (5,4) | 3.1003 | 1.440 | 119.562 | 122.362 | 113.893 | -7.0101 |
| (10,8) | 6.1705 | 1.433 | 119.473 | 124.333 | 115.020 | -7.0883 |
| (15,12) | 9.2493 | 1.432 | 119.456 | 124.707 | 115.232 | -7.0995 |
| (20,16) | 12.332 | 1.432 | 119.451 | 124.838 | 115.307 | -7.1032 |
| (25,20) | 15.415 | 1.432 | 119.448 | 124.898 | 115.341 | -7.1048 |
| (200,160) | 123.324 | 1.432 | 119.443 | 125.006 | 115.401 | -7.1076 |

In order to analyze the effect of chirality, we choose five tubes having approximately same radius but different chirality. We have plotted the normalized bond length with the chiral angle in Fig. 2. A maximum value of the bond length is obtained for chiral tubes of critical chiral angle equal to 13.892°. Above and below this critical chiral angle the bond length decreases.

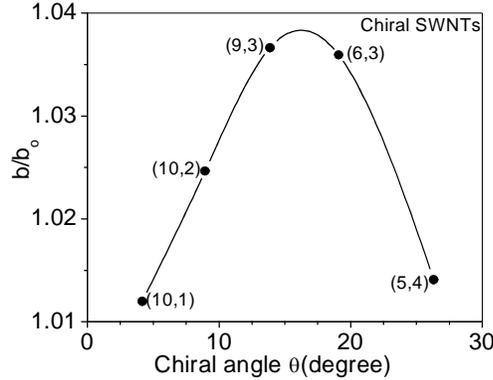

Figure 2: Normalized bond length versus chiral angle (degrees) for nearly same radius tubes of different chirality.

It is possible to explain the effect of the chirality on the bond length in the structure of chiral tubes. From Table I, the value of the bond lengths are found to be equal to 1.437Å, 1.455Å, 1.472Å, 1.479Å and 1.433Å for (10,1), (10,2), (9,3), (10,5) and (10,8) tubes, respectively, having approximately the same radii with different chirality. The smaller values of the bond length lie near the chirality of armchair tubes (chirality=1.0) and zigzag tubes (chirality=0.0). These results of the bond length disagree with calculations obtained by Jiang et al.[9]. They found three unequal bond lengths in the structure of chiral (4,2) tube. The values of these bond lengths obtained by them are equal to 1.460 Å, 1.467 Å and 1.455 Å and for (9,3) tube, those bond lengths become 1.453 Å , 1.454 Å and 1.451 Å. It should be noticed that the calculations of Ref. [9] did not reach a satisfactory graphite sheet bond length from the beginning. From our model, we were able to reproduce the bond length of 1.42 Å for graphite which agreed quite well with the experimental value [30] while Ref.[9] obtained 1.4507 Å which deviates largely from the experimental value. Three unequal bond angles are found in the structure of chiral tubes. Two of them, $\alpha$ and $\beta$, have values smaller than the ideal value while the third one $\beta$ found has a value larger than the ideal value (see Table I). The larger bond angle $\beta$ increases with increase in the tube radius. A maximum value of this bond angle is found for the tubes with chiral angle 13.892 Å (see Fig. 3). As regards the other two bond angles $\alpha$ and $\gamma$, the bond angle $\gamma$ increases with increase in the tube radius while the bond angle $\alpha$ decreases. The sum of these three bond angles is always smaller than $360^o$ for all tubes. This sum increases with increase in the tube radius and it is found to be independent of the chirality. Jiang et al.[9] have also found three unequal bond angles for chiral tubes. They obtained the value of $\beta$ larger than the ideal value and the other two bond angles are smaller. For (4,2) tube, they obtained the bond angles $\alpha$, $\beta$ and $\gamma$ equal to $114.19^o$, $120.59^o$ and $116.92^o$, respectively. These bond angles become $118.56^o$, $120.62^o$ and $119.36^o$ for (9,3) tube[9].

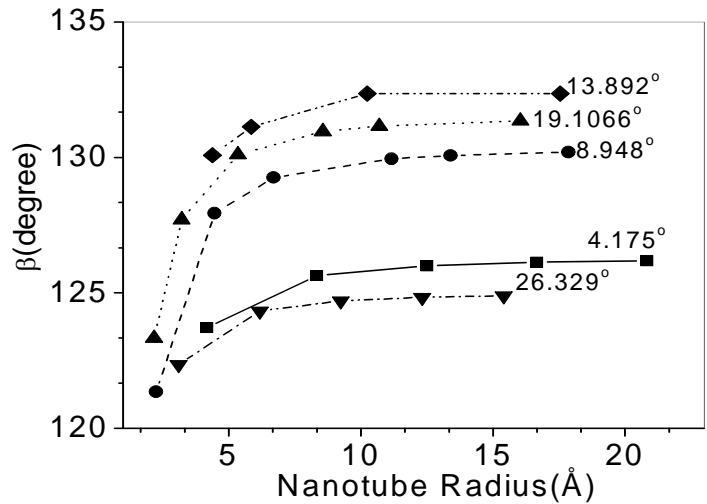

Figure 3: Bond angle $\beta$ as a function of the tube radius of five chiral angles. Smaller value of this angle also lies near the chiral angle of zigzag and armchair tubes and maximum value in mid region between them

The curvature energy of chiral tubes is found to be strongly dependent on chirality rather than the tube radius (see Table I for energy). Again, maximum effect is observed for tubes having chiral angle 13.892° (see Fig. 4). Far from this chiral angle, towards chirality of zigzag and armchair tubes, the curvature energy becomes lower but still has appreciably large value as compared to achiral tubes. Even large radius chiral tubes have a higher curvature effect compared with armchair and zigzag tubes [10,11]. For (100,10) tube with radius 83.35338Å curvature energy equals 0.316 eV/atom, close to a (10,1) tube of same chirality with radius 4.173486Å whose curvature energy equals 0.318 eV/atom. This result of the curvature effect reveals that the curvature effect in chiral tubes strongly depends only on the chirality. As the tube radius increases, the curvature effect slightly decreases (see Table I for the energy). A similar radius (around 80Å) for achiral tube would have insignificant curvature energy.

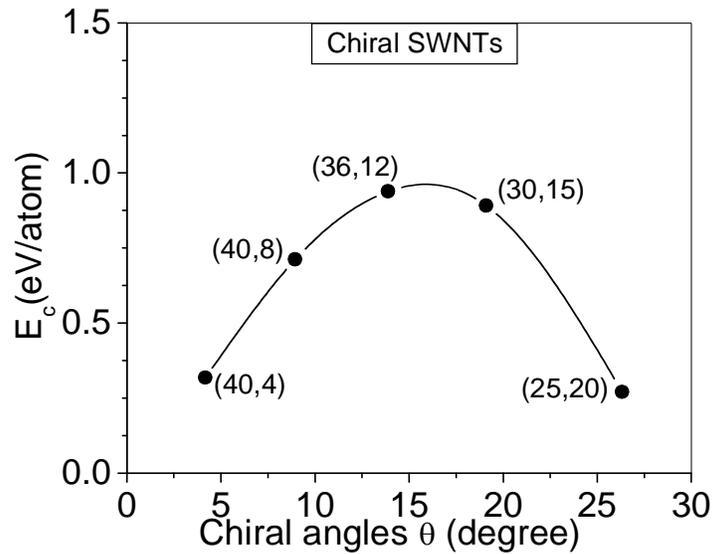

Figure 4: Curvature energy versus chiral angle for same radius tubes with different chirality.

**(B) Effect of length tube on the structure**

In contrast with the results of the effect of the tube length on the structural parameters in armchair tubes [10] and zigzag tubes [11], the results of length effect on the C-C bonds indicate that practically no effect of the tube length on the bond length and bond angles in the structure of chiral tubes takes place as we observe in Fig. 5(a) and Fig. 5(b), respectively. In these figures, we plotted the bond length and bond angles (for (10,1) tube) with length to radius ratio for chiral tubes of different chirality. There are similar curves for other tubes. The reason of this behavior is that the unit cell in chiral tubes is long enough to cancel any effect of the actual length as minimum length considered is multiple of these unit lengths.

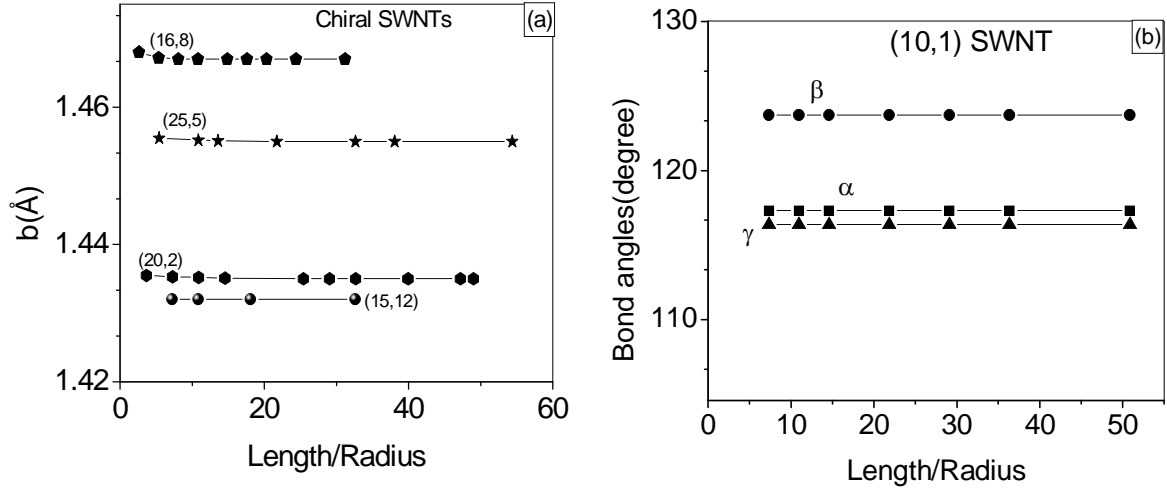

Figure 5: (a) Bond length as a function of the length to radius ratio for chiral (n, m) SWNTs at different chiral angles and (b) bond angles for chiral (10,1) SWNT.

**(C) Pressure effect**

In order to calculate the cross sectional structure under pressure for chiral tubes, we first assume the cross section to be circular and obtain minimum energy and bond lengths at various pressures. Subsequently, we allow change of the cross section to elliptical shape and recalculate structure. It must be noticed that even for bulk graphite, a high pressure study has shown that for pressure higher than 17 GPa, graphite undergoes a phase transition with bonding changes and bridging carbon atoms between graphite layers by converting half of the $\pi$-bonds between graphite layers to $\sigma$-bonds, indicating that interesting observations can be expected in the structure of carbon nanotubes under pressure[31].

**(1) Circular cross section**

To investigate the behaivour of the bond length and bond angles under hydrostatic pressure of chiral tubes, we choose four chirality tubes having chiral angles 4.175004°, 8.94876°, 19.1066° and 26.3295°. We also choose three tubes in each variety having different radii. Fig. 6 shows the variation of the bond length with applied pressure for chiral tubes. The bond length compresses under pressure, compression of this bond length is found to depend on the tube radius and chirality. The larger radii tubes of same chirality are easily compressed as compared to smaller radii tubes. As can be noticed from Fig. 6, the bond length reduction at pressure equal to 10 GPa from its value at zero pressure is around 2.5% and 9.8% in (10,1) and (40,4) tubes, respectively. As far chirality effect on compression, at 10 GPa, this bond length reduction is close to 9.6% and 9.8% in (40,4) and (30,15) tubes of approximately the same radius, respectively.

The required pressure to compress the bond length to graphitic bond length value $b_o$ is found to depend on the chirality and tube radius. This required pressure is denoted by $P_c^{(n,m)}$. A plot of $P_c^{(n,m)}$ with chiral angle has been given in Fig. 6(e). All tubes in each curve have approximately the same radii but with different chirality. We have also presented these data in Table II. We observe that all tubes of chiral angle equal to 19.948° require highest pressure to compress them to graphitic bond length value. The range of $P_c^{(n,m)}$ varies approximately from 2 to 10 GPa. These important results indicate that the rigidity of a chiral single-walled nanotube depends on both the radius and chirality of the tube. Bond angles under applied pressure remain have the same values as at ambient pressure for all chiral tubes (Fig. 6(f)). A better estimate about the rigidity of the tubes is provided by the results of the bulk modulus. The results of bulk modulii have been listed in Table II. In this table, we see that the value of the bulk modulus depends of the tube radius and the chirality. The small radius tubes have larger value of bulk modulus. However, a careful look at bulk modulus of tubes (10,5) and (10,8) reveals that although (10,8) tubes is bigger in radius, it has larger bulk modulus. This is because of chirality effect. Close to critical chiral angle as is in case of (10,5), the chirality seems to reduce the bulk modulus.

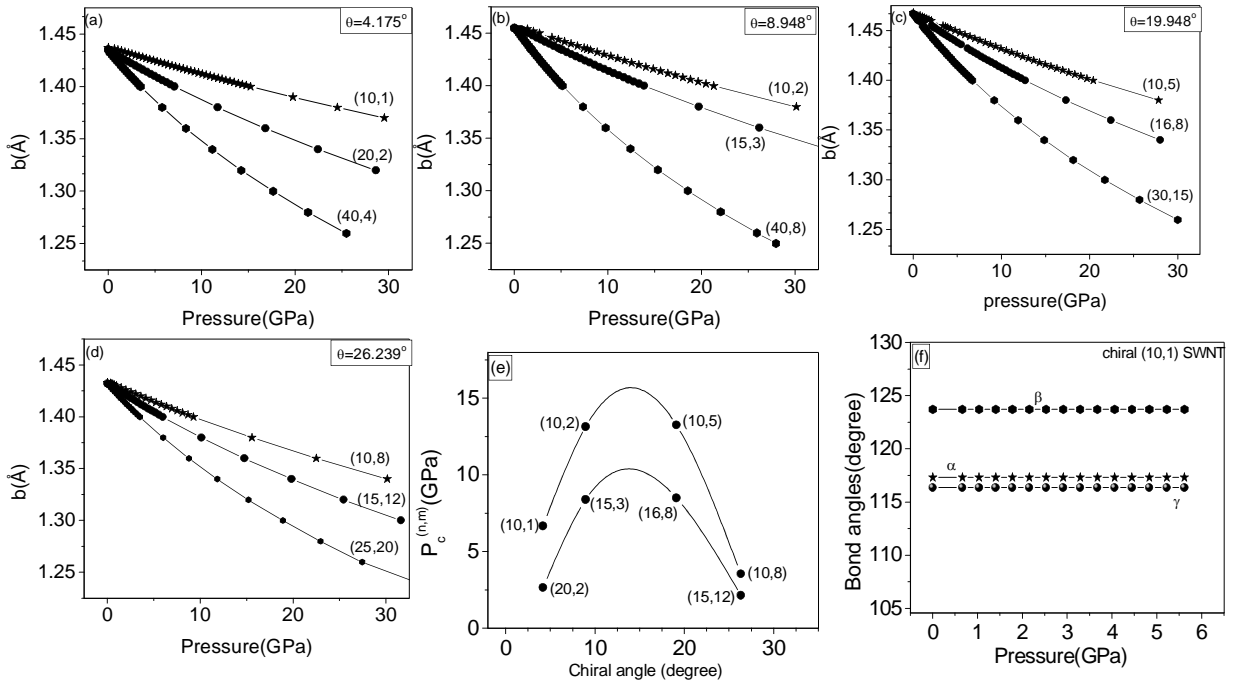

Figure 6: (a), (b), (c), and (d) bond length as a function of the applied pressure for different radii tubes at different chiral angles. (e) Pressure, $P_c^{(n,m)}$, to compress bond length to 1.42Å as a function of chiral angle for the same radii tubes with different chirality and (f) bond angles with applied pressure plotted only for (10,1) tube.

Table II: Bulk modulus (GPa) at zero pressure at different chiral angles for chiral tubes.

| Chiral angle=4.175° | | | Chiral angle=8.948° | | |
|---|---|---|---|---|---|
| SWNT | Radius(Å) | Bulk moduls | SWNT | Radius(Å) | Bulk moduls |
| (10,1) | 4.17348 | 131.5175 | (10,2) | 4.46636 | 113.4222 |
| (20,2) | 8.33593 | 66.12793 | (15,3) | 6.69936 | 75.47363 |
| (40,4) | 16.6707 | 32.59589 | (40,8) | 17.8654 | 28.31979 |
| Chiral angle=19.1066° | | | Chiral angle=26.3295° | | |
| (10,5) | 5.35335 | 85.60864 | (10,8) | 6.170521 | 91.96869 |
| (16,8) | 8.55952 | 54.52480 | (15,12) | 9.249323 | 63.76404 |
| (30,15) | 16.0491 | 28.96707 | (25,20) | 15.41015 | 36.34271 |

**(2) Shape transition**

We also examine the existence of the shape transition, from circular to oval cross section, under pressure for chiral tubes. We study the effect of the tube radius and chirality on the transition pressure and the behavior of the bond length at this transition. For these goals, we choose chiral tubes having different radii of the same chirality and tubes having approximately the same radii of different chirality. In Fig. 7, we have plotted the variation of energy as a function of applied pressure for two chiral tubes (10,1) and (20,2) having different radius of the same chirality (i.e., chiral angle 4.175004°). The curve contains two parts: AB curve represents a circular cross section while CD curve represents an elliptical cross section. Each point on the CD curve corresponds to different value of $b_e/a_e$, where the $b_e$ and $a_e$ are the shorter and longer axes in elliptical cross section. We observe that at point C the energy of tubes with elliptical cross section is lower than that energy with circular cross section. This means that the tube at transition pressure (point C) begins to appear elliptical in cross section. Below the value of transition pressure, the energy with circular cross section is slightly lower. The value of elliptical aspect ratio at point C is found equal to 0.999 for all chiral tubes with different radii and different chirality.

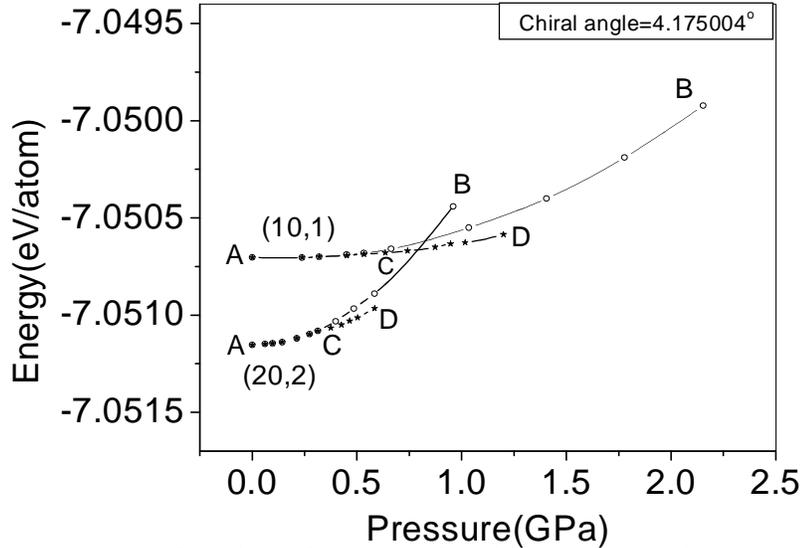

Figure 7: Energy as a function of pressure for chiral (10,1) and (20,2) tubes. AB and CD curves are corresponding to circular and elliptical cross section, respectively. At point C in both curves, $b_e/a_e$=0.999.

In Table III, we present the results of the transition pressure for chiral tubes with different radii and chirality. We have also plotted the transition pressure as a function of the tube radius of the same chirality in Fig. 8. We observe that the transition pressure depends on the tube radius. As the tube radius increases the required pressure to occur this transition in the cross section decreases.

Our calculations of the transition pressure for chiral tubes indicate that the chirality of tubes also affects the value of transition pressure. To show this effect of the chirality on the transition pressure, we compare the results of two tubes having approximately the same radii and different chirality, chiral tubes (15,3) and (10,8). They have radius close to 6 Å as can be seen in Table I. The value of transition pressure for these tubes is found to be equal to 1.667 GPa and 0.362 GPa, respectively which differ significantly. Infact, it seems that the chiral tubes midway between zigzag and armchair tubes would be difficult to bring in shape change. These results are consistent with all other effects of chiral tube behavior. From these results, it is very clear to conclude that the transition pressure also depends on the chirality in addition to the tube radius. Our results of the chirality dependence of the transition pressure disagree with the results observed by Eilliot et al.[16]. They observed that the transition pressure depends on the diameter of nanotubes and not on its chirality in a bundle or rope of single-walled nanotubes. In fact, detailed pressure dependent measurement of phonon in carbon nanotubes through Raman spectroscopy manifests this change of the cross section through disappearing of RBM [18,32].

The results of bond length at transition pressure for chiral tubes shows very interesting behavior indicating that tubes with different radii and different chirality have approximately the same value of the bond length (see Fig 9.).

Table III: Chiral angle($\theta$), longer radius ($a_e$), shorter radius ($b_e$), transition pressure ($P_T$) and bond length ($b$) for chiral tubes in five chiral angles with elliptical aspect ratio equal to 0.999.

| $\theta$ (degree) | (n,m) | $a_e$ (Å) | $b_e$ (Å) | $P_T$ (GPa) |
|---|---|---|---|---|
| 4.175 | (10,1) | 4.167 | 4.163 | 0.635 |
| 4.175 | (20,2) | 8.298 | 8.306 | 0.375 |
| 4.175 | (30,3) | 12.41 | 12.40 | 0.251 |
| 8.949 | (15,3) | 6.561 | 6.554 | 1.667 |
| 13.892 | (12,4) | 5.705 | 5.699 | 1.853 |
| 19.106 | (10,5) | 5.240 | 5.235 | 1.454 |
| 26.329 | (10,8) | 6.157 | 6.151 | 0.362 |

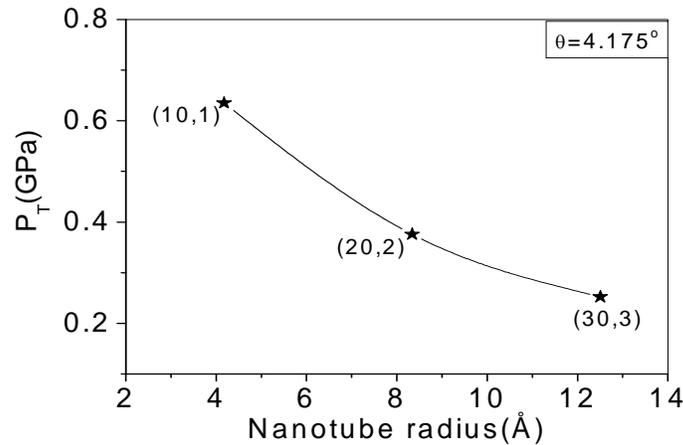

Figure 8: Transition pressure as a function of the tube radius with chiral angle 4.175°. There is a radius dependance of the first transition pressure.

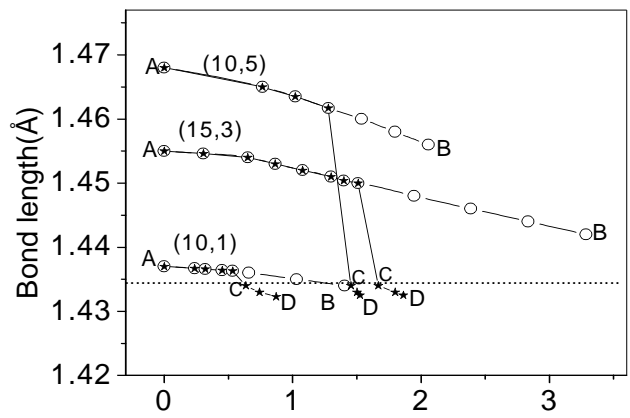

Figure 9: Bond length as a function of pressure for chiral (10,1) , (20,2) and (10,5) tubes. AB and CD curves are corresponding to circular and elliptical cross section, respectively. At transition pressure, all chiral tubes have approximately the same value of the bond length.

### IV- Summary and Conclusion

In this paper, we investigate the effect of radius, length and chirality of the tube on bond lengths and bond angles of the uncapped chiral (n,m) SWNTs. For this goal, we construct the SWNT through different bond lengths using helical and rotational symmetries. We ascertained that only one bond length characterizes chiral tubes in contrast to achiral tubes which are characterized by two bond lengths. This result was obtained on the basis of the energy calculation which was found to be minimum only for equal bond lengths .

This bond length is found to strongly depend on the chirality and slightly on the tube radius. For all chiral tubes, the value of this bond length is found to be always larger than that in the graphitic value. One of three bond angles is found to be larger than that the ideal value. The values of other two unequal bond angles are found smaller than that in graphite. The curvature effect is dominated by the effect of chirality in addition to the effect of the tube radius. Maximum curvature effect occurs at a critical chiral angle 13.899°, which is in the middle range of chirality when one considers the zigzag on the one end and armchair on the other extreme of chirality. Above and below this critical chiral angle the curvature effect decreases. In contrast with achiral tubes, the bond length and bond angles are found to remain practically independent of the tube length.

We also calculated the structure of chiral SWNTs under hydrostatic pressure assuming the cross section has a circular shape. The three equal bond lengths under pressure compress by the same way and at some pressure $P_c^{(n,m)}$ these reach the bond length to graphitic value. $P_c^{(n,m)}$ at which graphitic bond length is obtained is found to depend on the chirality and tube radius. $P_c^{(n,m)}$ is largest for critical chiral angle tubes. Bond angles under pressure remain unaltered. Results of the bulk modulus indicate that the rigidity of chiral tubes also depends on the chirality in addition of the tube radius. Again, critical angle chiral tubes, show reduced value of bulk modulus. We also found that there is an existence of the shape transition, transition from circular to oval cross section for chiral tubes. There is a radius and chirality dependence of the transition pressure . However, it is interesting to observe that the bond length  at this shape transition is practically constant(~1.434A) for all chirality and radius tubes.

We believe that the results of this work and our previous work on armchair and zigzag tubes reveal the difference between the structures of three types of single-walled carbon nanotubes at ambient pressure.  Under hydrostatic pressure, several new observations in relation to bond lengths and chirality  have been made which differentiate the chiral tubes from achiral ones These results can be used to distinguish between these three types of SWNTs. As a result of this work  we  hope that a complete  picture of SWNTs on the behavior of the structural parameters at ambient pressure and under hydrostatic pressure is now available.